%% file: UCN2023.tex
\documentclass[aps,prc,twocolumn,superscriptaddress]{revtex4-2} 
\pdfoutput=1 
\usepackage[english]{babel} 

\usepackage{newtxtext}
\usepackage[upint]{newtxmath}
\usepackage{bm}
\usepackage{graphicx}
\usepackage{tikz}
\usepackage{hyperref}

\begin{document}

\title{Improving of ultracold neutron traps coated with liquid helium using capillarity and electric field}

\author{Pavel\,D.~Grigoriev}
\email{grigorev@itp.ac.ru}
\affiliation{L.\,D.~Landau Institute for Theoretical Physics, 142432, Chernogolovka, Russia}
\affiliation{National University of Science and Technology ''MISiS'', 119049, Moscow, Russia} 
\author{Arseniy\,V.~Sadovnikov}
\affiliation{Lomonosov Moscow State University, 119991, Russia}
\affiliation{I.\,M. Sechenov First Moscow State Medical University, 119991, Russia}
\author{Vladislav\,D.~Kochev}
\affiliation{National University of Science and Technology ''MISiS'', 119049, Moscow, Russia} 
\author{Alexander\,M.~Dyugaev}
\affiliation{L.\,D.~Landau Institute for Theoretical Physics, 142432, Chernogolovka, Russia}

\begin{abstract}
To increase the storage time of ultracold neutrons (UCN) inside the material traps it is promising to cover the trap walls by liquid $^4$He, the material which does not absorb neutrons at all. A rough side wall of UCN trap holds the required amount of $^4$He by the capillary effects, but the edges of wall roughness remain insufficiently coated.
Here we propose to apply an electric voltage to these rough side walls of UCN traps to increases the thickness of liquid He on the wall edges and to cover the entire wall surface by sufficiently thick helium films. This completely protects UCN from being absorbed inside the trap walls. We estimate the required electric field and voltage for several possible designs of UCN traps. This improvement may give rise to a new generation of ultracold neutron traps with very long storage time. We also estimate the influence of this electric field on the dispersion of ripplons -- the quanta surface waves, which give the main contribution to the inelastic UCN scattering at low temperature.  
\end{abstract}

\keywords{Ultracold neutrons, neutron lifetime measurement, liquid helium}
\maketitle

\section{Introduction}


The precise measurements of neutron lifetime $\tau _{n}$ are important for
elementary particle physics, astrophysics and cosmology (see \cite%
{Abele/2008,Musolf/2008,Dubbers/2011,WietfeldtColloquiumRMP2011,GONZALEZALONSO2019165}
for reviews). The Big-Bang nucleosynthesis and chemical element formation
depends on $\tau _{n}$. In combination with spin-electron asymmetry measured
in polarized-neutron decay experiments \cite%
{PhysRevLett.105.181803,Beam2019PhysRevLett.122.242501,PhysRevC.101.035503}
the $\tau _{n}$ measurements give both vector and axial coupling constants
of weak interaction in hadronic current between nucleons, which differs from
the quark current due to a renormalization by strong interaction. The search
for a non-zero electric dipole moment (EDM) of neutrons\cite%
{Pospelov2005,Baker/2006,SerebrovJETPLetters2014} impose the limits on CP
violation. The resonant transions between discrete quantum energy levels of
neutrons in the earth gravitational field \cite%
{NesvizhevskyNature2002,UCNResonancePhysRevLett.112.151105} probe the
gravitational field on a micron length scale and impose constraints on dark
matter.

The ultracold neutrons (UCN) with energy lower than the neutron optical
potential of typical materials, i.e.\ $\lesssim 300$ neV, are widely
employed in neutron experiments \cite%
{Golub/1991,Ignatovich/1990,Ignatovich1996,PhysRevLett.105.181803,PhysRevC.101.035503,Baker/2006,SerebrovJETPLetters2014,NesvizhevskyNature2002,UCNResonancePhysRevLett.112.151105,Serebrov2008PhysRevC.78.035505,ArzumanovPhysLettB2015,Serebrov2017,Serebrov2018PhysRevC.97.055503,Review2019Pattie}%
. These UCN can be trapped for many minutes in specially designed ''neutron
bottles'' \cite%
{Serebrov2008PhysRevC.78.035505,ArzumanovPhysLettB2015,Serebrov2017,Serebrov2018PhysRevC.97.055503,Review2019Pattie}%
, where the earth gravitational field $100$ neV per meter plays an important
role in UCN storage and manipulation \cite%
{Golub/1991,Ignatovich/1990,Ignatovich1996,Serebrov2008PhysRevC.78.035505,ArzumanovPhysLettB2015,Serebrov2017,Serebrov2018PhysRevC.97.055503,Review2019Pattie}%
. The Fomblin grease is currently used to cover the UCN trap walls \cite%
{Serebrov2008PhysRevC.78.035505,ArzumanovPhysLettB2015,Serebrov2017,Serebrov2018PhysRevC.97.055503,Review2019Pattie,MAMBOPhysRevLett.63.593,PICHLMAIER2010}
in the bottle UCN experiments and allows reaching the highest accuracy of
neutron lifetime measurements: $\tau_{n}$ = 881.5 $\pm$ 0.7(stat) $\pm$~0.6(syst) s \cite{Serebrov2018PhysRevC.97.055503}. 
Because of of the neutron magnetic
moment of $60$ neV/T, magneto-gravitational trapping of UCN is feasible and
promising too \cite%
{Huffman2000,PhysRevC.94.045502,PhysRevC.95.035502,Ezhov2018,Pattie2018},
giving a comparable claimed accuracy. However, the non-uniformity of
magnetic field produces considerable losses of spin-polarized UCN in such
magnetic traps, and an accurate estimate of these losses to account them for
is a difficult problem. Therefore, in spite of high claimed precision of
magnetic-trap $\tau _{n}$ measurements, the corresponding values \cite%
{Ezhov2018} $\tau_{n}\approx$ 878.3 $\pm$ 1.6(stat) $\pm$ 1.0(syst) s or \cite%
{Pattie2018} $\tau_{n}\approx$ 877.7 $\pm$ 0.7(stat)$^{+0.4}_{-0.2}$(syst) s are about 
4 s smaller than in the material-bottle UCN experiments.

The main alternative to using UCN in neutron lifetime measurements is the
cold neutron beam, giving $\tau_{n}$ = 887.7 $\pm$ 1.2(stat) $\pm$ 1.9(syst) s 
\cite{BeamPhysRevC.71.055502,BeamPhysRevLett.111.222501,BeamReview2020}. The
discrepancy between $\tau _{n}$ measured by beam and UCN material- or
magnetic-trap methods is beyond the estimated errors. This
\textquotedblright neutron lifetime puzzle\textquotedblright\ is a subject
of extensive discussion till now \cite%
{BeamReview2020,DarkMatter2021PhysRevD.103.035014,Serebrov2021PhysRevD.103.074010,Serebrov2019}%
. Presumably, it is due to systematic errors in beam experiments \cite%
{Serebrov2021PhysRevD.103.074010}, but unconsidered UCN losses in bottle $%
\tau _{n}$ measurements are not excluded yet. 
As has been shown by analyzing the neutron $\beta$-decay asymmetry \cite{Dubbers2019}, 
it is unlikely that this discrepancy is caused by other new physics like 
additional neutron decay channels or dark matter \cite{BeamReview2020,DarkMatter2021PhysRevD.103.035014}.
 Hence, reducing the UCN losses in material traps is 
crucial for various neutron experiments.

The precision of current neutron lifetime measurements using UCN traps, both
material and magnetic, is limited by the accuracy of estimating of neutron
escape rate from the traps, which is the main source of systematic errors \cite%
{Golub/1991,Ignatovich/1990,Ignatovich1996,Goremychkin2017,Serebrov2018PhysRevC.97.055503,Review2019Pattie}.
At present, material UCN traps coated with Fomblin grease provided the
highest accuracy of $\tau _{n}$ measurements. Any collision of a neutron
with trap wall leads to $\sim 10^{-5}$ probability of neutron absorption by
the wall material \cite%
{Golub/1991,Ignatovich/1990,Ignatovich1996,WietfeldtColloquiumRMP2011}. The
neutron lifetime $\tau _{n}$ is estimated by extrapolation of the measured
lifetime $\tau _{1}$ of neutrons stored in the trap to the zero neutron
losses by a careful variation of the bottle geometry and/or temperature, so
that the loss contribution from trap walls can be accounted for. The
extrapolation interval is rather large, usually, $\tau _{n}-\tau _{1}\gtrsim
30$ s, which limits the precision of $\tau _{n}$ measurements, because
estimating the systematic error with accuracy better than $5$\% is a very
hard problem. This estimate is complicated by the dependence of UCN absorption 
probability on the angle of incidence during each collision. The usually applied
assumption \cite{Serebrov2018PhysRevC.97.055503} of the uniform distribution 
of neutron velocity direction with respect to the trap surface is violated 
for the collisions with side walls because the vertical UCN velocity 
component depends on the height above trap bottom due to gravity. This difficulty
can be overcome by Monte-Carlo similations of UCN losses taking into account 
the calculated incidence angles of each collision for the particular trap geometry 
provided the initial momentum distribution of UCN is known. A more serious problem is 
the surface roughness, which makes impossible the exact calculation of UCN 
loss probability during each collision. Hence, the accuracy of the estimates 
of UCN loss rate in material traps cannot be strongly improved.

The surface-to-volume ratio in material traps and the UCN
losses on trap walls can be reduced by increasing the trap size. Note that
the UCN material traps covered by Fomblin oil must be kept at a low
temperature $T<90$~K to reduce the inelastic neutron scattering. In most
precise recent $\tau _{n}$ measurements \cite{Serebrov2018PhysRevC.97.055503}
the extrapolation interval $\tau _{n}-\tau _{1}$ was reduced to only $20$
seconds by increasing the size of high-vacuum UCN material trap to $2$~m,
making the dimensions of external vacuum vessel $4.2$~m. However, a further
increase in the size of UCN traps seems technically problematic and not very
useful, because main neutron losses come from their collisions with
trap bottom rather with its side walls. The rate of neutron collisions with the
trap bottom is determined by the UCN kinetic energy along the $z$-axis and 
does not depend on the trap size. Hence, the precision of $\tau_n$-measurements 
in traditional UCN traps seems to reach its limit.

A possible qualitative step to further reduce the neutron escape rate from
UCN traps is to cover the trap walls by liquid $^{4}$He, the only material
that does not absorb neutrons \cite{Golub1983,Bokun/1984,Alfimenkov/2009}.
However, $^{4}$He provides a very small optical potential barrier $%
V_{0}^{\mathrm{He}}=18.5\ \mathrm{neV}$ for the neutrons, which is much smaller than
the barrier height $V_{0}^{\mathrm{F}}\approx$ 106~neV of Fomblin oil. Only
UCN with kinetic energy $E<V_{0}^{\mathrm{He}}$ can be effectively stored in such a
trap. The corresponding maximum height of UCN $h_{\max
}=V_{0}^{\mathrm{He}}/m_{n}g\approx 18$~cm, where $m_{n}=1.675\times 10^{-24}$~g is the
neutron mass. The UCN phase volume and their density in the He trap is
reduced by the factor $(V_{0}^{\mathrm{F}}/V_{0}^{\mathrm{He}})^{3/2}\approx 13.7$ as compared
to the Fomblin coating. This raises the statistical errors. However, the UCN
density increases as technology develops \cite%
{PhysRevC.99.025503,ZimmerPhysRevC.93.035503}, and this reduction of neutron
density may become less important than the advantage from a decrease of UCN
loss rate.

The second problem with the liquid $^{4}$He coating of UCN trap walls is a very low
temperature $T<0.5$~K. At higher temperature $^{4}$He vapor inelastically
scatters UCN, giving them energy $\sim k_{B}T\gg V_{0}^{\mathrm{He}}$. At $T<0.5$~K
the concentration of $^{4}$He vapor $n_{V}\propto \exp \left( -7.17/T\left[\mathrm{K}\right] \right) $ is negligibly small. Another source of inelastic UCN
scattering are ripplons, the thermally activated quanta of surface waves.
They lead to a linear temperature dependence of scattering rate \cite%
{Grigoriev2016Aug}, surviving even at ultralow temperature. However, the
strength of neutron-ripplon interaction is rather small \cite%
{Grigoriev2016Aug}, which makes feasible the UCN storage in He-covered
traps. Moreover, the linear temperature dependence of UCN losses due to
their scattering by ripplons is very convenient for taking into account this
systematic error.

The third problem with liquid $^{4}$He is too thin helium film covering the
side walls of the trap. $^{4}$He is superfluid below $T_{\lambda }=2.17$~K
and covers not only the floor but also the walls and the ceiling of the trap
because of the van-der-Waals attraction. On flat vertical walls few
centimeters above the He level, the thickness of helium film is expected to
be only $d_{\mathrm{He}}^{\min }\approx 10$~nm$<\kappa _{0}^{-1}$, while the neutron
penetration depth into the liquid helium is $\kappa _{0\mathrm{He}}^{-1}=\hbar /\sqrt{%
2m_{n}V_{0}^{\mathrm{He}}}\approx 33.3\ \mathrm{nm}>d_{\mathrm{He}}$. Hence, the tunneling
exponent%
\begin{equation}
\psi \left( 0\right) /\psi \left( d_{\mathrm{He}}\right) \sim \exp \left( -\kappa
_{0\mathrm{He}}d_{\mathrm{He}}\right) ,  \label{psiS1}
\end{equation}%
of the neutron wave function $\psi $ inside He is not sufficient to strongly
reduce the neutron losses on the trap walls. A more accurate calculation of
the neutron wave function near a solid wall covered with liquid helium \cite%
{GrigorievPRC2021} increases the estimate of $\psi \left( 0\right) $ by $%
\sim 30\%$ as compared to Eq. (\ref{psiS1}) for relevant UCN kinetic energy $%
E<0.8V_{0}^{\mathrm{He}}$, making the problem of too thin $^{4}$He film even more
serious. The required thickness of helium film for a good protection of UCN
is $d\leq d_{\mathrm{He}}^{\ast }=100~\mathrm{nm}$ \cite{GrigorievPRC2021}.
An idea \cite{Bokun/1984} of using a rotating He vessel for UCN storage to
increase He thickness on side walls has a drawback that the rotating liquid
generates additional bulk and surface excitations, leading to inelastic
neutron scattering. Therefore, one needs a time-independent covering of the
trap walls by liquid $^{4}$He. A possible solution of this problem, proposed
recently \cite{GrigorievPRC2021,Grigoriev2021}, is based on using a rough
surface of trap side walls. This rough surface holds liquid helium of
sufficient thickness by the capillary effect, if the wall roughness has much
smaller period $l_{R}$ than the $^{4}$He capillary length $a_{\mathrm{He}}=\sqrt{%
\sigma _{\mathrm{He}}/g\rho _{\mathrm{He}}}=0.5$~mm, where $\sigma _{\mathrm{He}}=0.354$~dyn/cm is the
surface tension coefficient of liquid $^{4}$He, $g=980$~cm/s$^{2}$ is the
free fall acceleration, and the liquid $^{4}$He density $\rho _{\mathrm{He}}\approx
0.145$~g/cm$^{3}$. The calculations showed \cite{GrigorievPRC2021,Grigoriev2021} 
that one needs even the smaller period of wall roughness 
$l_{R}\lesssim 4a_{\mathrm{He}}^2 /h$ to hold superfluid $^{4}$He on the heigh $h$
above the helium level.

In Ref. \cite{Grigoriev2021} it was argued that a simple triangular wall
roughness of period $l_{R}\lesssim 5$~\textmu m, as in the mass-produced diffraction
gratings \footnote{Diffraction gratings with the period 
$l_{\mathrm{R}}=1$~\textmu m and dimensions $1.524\times 0.1524\,\mathrm{m}$
are available at a price of $\$20$ at www.amazon.com.}, is better than rectangular roughness to hold a shielding helium film
in UCN traps. The diffraction gratings with the period $l_{\mathrm{R}%
}\approx 4$~\textmu m and depth $h_{\mathrm{R}}\approx 0.2$~\textmu m are already actively used for the scattering of UCNs \cite%
{Kulin2016,Kulin2019}. 
However, the thickness of He film on sharp peaks of this
triangular roughness remains less than $\kappa _{0\mathrm{He}}^{-1}$, which leaves $%
\lesssim 5\%$ of wall surface insufficiently coated. In this paper we
propose to use an electrostatic potential to increase the efficiency of such
helium-covered UCN traps and to coat the remaining unshielded surface area. 
This may completely eliminate the UCN losses from the absorption inside the trap 
walls and start a new generation of ultracold neutron traps with very long storage time.  

\section{Energy functional of liquid helium and required field strength}

To describe the profile of the helium film on the rough surface, it is
necessary to minimize the energy functional of this film 
\begin{equation}
E_{\mathrm{tot}}=V_{\mathrm{g}}+E_{\mathrm{s}}+V_{\mathrm{w}}+V_{\mathrm{e}}.
\label{eq:EtotM}
\end{equation}%
This functional (\ref{eq:EtotM}) differs from that considered in Refs. \cite%
{GrigorievPRC2021,Grigoriev2021} by the new term $V_{\mathrm{e}}$, which
comes from the polarization energy of helium in a non-uniform electric
field, as described below. The first three terms in Eq. (\ref{eq:EtotM}) are
the same as in Refs. \cite{GrigorievPRC2021,Grigoriev2021}. $V_{\mathrm{g}}$
is the gravity term given by the expression 
\begin{equation}
V_{\mathrm{g}}=\rho _{\mathrm{He}}g\int\!\mathrm{d}^{2}\bm{r}_{\Vert } zd_{\mathrm{He}}\left( \bm{r}%
_{\Vert }\right) ,  \label{eq:Vg}
\end{equation}%
where $\bm{r}_{\Vert }=\{x,\,z\}$ is the two-dimensional vector of the
horizontal, $x$, and vertical, $z$, coordinates on the wall, 
\begin{equation}
d_{\mathrm{He}}\left( \bm{r}_{\Vert }\right) =\xi \left( \bm{r}%
_{\Vert }\right) -\xi _{\mathrm{W}}\left( \bm{r}_{\mathrm{W}}\right)
\label{eq:dHeDef}
\end{equation}%
is the thickness of the helium film. The functions $\xi (\bm{r}_{||})$
and $\xi _{\mathrm{W}}\left( \bm{r}_{\Vert }\right) $ describe the
profiles of the He surface and of the solid trap wall.

The second term $E_{\mathrm{s}}$ in Eq. (\ref{eq:EtotM}) describes the
surface tension energy and is given by the formula 
\begin{equation}
E_{\mathrm{s}}=\sigma _{\mathrm{He}}\int \!\mathrm{d}^{2}
\bm{r}_{\Vert } \left( \sqrt{1+\left[ 
\nabla \xi \left( \bm{r}_{\Vert }\right) \right] ^{2}}-1\right) .  \label{eq:Es}
\end{equation}%
Here we subtracted the constant term of a flat surface $E_{\mathrm{s0}%
}=\sigma _{\mathrm{He}}\int \!\mathrm{d}^{2}\bm{r}_{\Vert }$.

The third van der Waals term $V_{W}$ in Eq. (\ref{eq:EtotM}) describes the
attraction of helium to the wall material. It is significant only at
small distance and leads to coating of the entire wall surface by a
superfluid helium film thicker than $d_{\mathrm{He}}^{\mathrm{min}}\approx
10$~nm . As shown in Refs. \cite{GrigorievPRC2021,Grigoriev2021}, the
capillary effects compensate the gravity term and hold much thicker helium
film on the height $h$ above helium level if the characteristic length scale
of the wall roughness does not exceed$\ l_{\mathrm{R}}^{\max }=4a_{\mathrm{He%
}}^{2}/h$. For the wall roughness of the shape of a triangular
grid, as proposed in Ref. \cite{Grigoriev2021}, $l_{\mathrm{R}}^{\max }$
gives its maximal period.

The electric energy term $E_{\mathrm{e}}$ in Eq. (\ref{eq:EtotM}),
describing the polarization energy of helium in a non-uniform electric
field, is%
\begin{equation}
E_{\mathrm{e}}=-\frac{\varepsilon _{\mathrm{He}}-1}{4\pi }\int \! \mathrm{d}^{3}\bm{r} E^{2}\left( \bm{r}%
\right),  \label{Ee}
\end{equation}%
where the integral in taken over the volume occupied by liquid He, and $%
\varepsilon _{\mathrm{He}}=1.054$ is the dielectric constant of $^{4}$He. On the trap
side wall Eq. (\ref{Ee}) rewites as 
\begin{equation}
E_{\mathrm{e}} = -\frac{\varepsilon _{\mathrm{He}}-1}{4\pi }\int \!\mathrm{d}^{2}\bm{r}_{\Vert } E^{2}\left( 
\bm{r}_{\Vert },d_{\mathrm{He}}\right) d_{\mathrm{He}}\left( \bm{r}%
_{\Vert }\right) .  \label{EeSW}
\end{equation}%
This term is new as compared to Refs. \cite{GrigorievPRC2021,Grigoriev2021}.

The surface profile of superfluid $^{4}$He corresponds to the constant
energy $E_{\mathrm{tot}}=E_{0}=\mathrm{const}$ in Eq. (\ref{eq:EtotM}) of a tiny
helium volume. This constant energy is the same, as for liquid $^{4}$He on
the surface above the bottom of UCN trap. Eqs. (\ref{eq:EtotM})-(\ref{Ee})
assume liquid helium to be incompressible, which is a rather good
approximation.\ To find the $^{4}$He surface profile on the side wall one
needs to know the spatial distribution of the absolute value of electric
field strength $E^{2}\left( \bm{r}\right) $.

The capillary effects do not help to cover the wall edges by a sufficiently
thick helium film. Eqs. (\ref{eq:EtotM}),(\ref{eq:Vg}) and (\ref{EeSW})
allow to estimate the required electric field strength $E_{\ast }(h)$ at the
surface of liquid helium on the side wall at height $h$ to hold the helium
film even without the capillary effects. The electric term (\ref{EeSW})
compensates the gravity term (\ref{eq:Vg}) if $V_{\mathrm{g}}+E_{\mathrm{e}%
}<0$, which gives the electric field strength 
\begin{equation}
E\geq E_{\ast }=\sqrt{4\pi \rho _{\mathrm{He}}gh  /\left( \varepsilon
_{He}-1\right) }.  \label{Emax}
\end{equation}%
For $h=h_{\max }=18~\mathrm{cm}$ this gives a target electric field $E_{\ast
}\approx 230$~\textrm{kV/cm}. Such a strong electric field appears because of
a weak $^{4}$He polarization, $\varepsilon _{\mathrm{He}}-1=0.054\ll 1$.
Nevertheless, it is still much smaller than the field of dielectric
breakdown $E_{\max }>1$~\textrm{MV/cm} of $^{4}$He \cite{Ito2016}. Therefore,
it is theoretically achievable.

Fortunately, one does not need to apply such a strong field $E_{\ast }$ at
the whole side-wall surface but only near the edges of its roughness. Near
these edges the electric field can be easily increased if the rough wall
itself serves as an electrode. Below we consider this in more detail for the
triangular wall roughness and show, that we need an external electric field $%
E_0$ several times weaker than $E_{\ast }$.

\section{Electric field strength and helium film thickness near a triangular edge}

\begin{figure}[htb]
	\begin{tikzpicture}[every node/.style={inner sep=0,outer sep=0}]
		\node (picture) {\includegraphics[width=\linewidth]{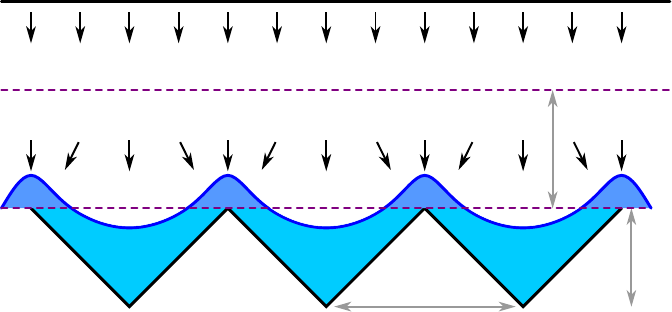}};
		\node[above=0.2cm,right=1.3cm] at (picture.north west) {electrode, $V=V_0$};
		\node[below=0.2cm,right=0.8cm] at (picture.south west) {grounded trap wall, $V=0$};
		\node[below=0.9cm,right=0cm] at (picture.north west) {$E\approx E_0\approx \operatorname{const}$};
		\node[below=1.5cm,right=0cm] at (picture.north west) {$E=E(\bm{r})$};
		\node[above=0.3cm,left=3cm] at (picture.south east) {$l_R$};
		\node[above=0.7cm,left=0.55cm] at (picture.south east) {$h_R$};
		\node[above=2.5cm,left=1.6cm] at (picture.south east) {$\sim \!l_R$};
	\end{tikzpicture}
	
	\caption{Triangular roughness of solid UCN trap wall coated
		with liquid helium with periodically modulated surface (blue solid line). 
		The applied electric field is the strongest near the wall edges, which 
		attracts $^4$He to these edges and coat them with a helium film of sufficient thickness.}
	\label{FigTriang}
\end{figure}
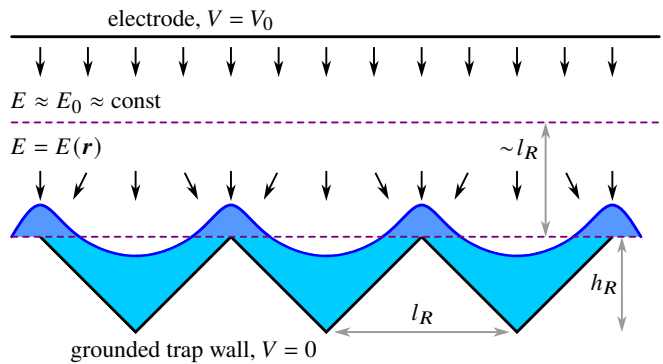

We consider a grounded metallic rough trap wall with voltage $V=0$ covered
with $^{4}$He. Another electrode at voltage $V_{0}$ is separated at some
characteristic distance $L$ (see Fig. \ref{FigTriang}). The electric potential $V\left(\bm{r}%
\right) $ raises from $V=0$ at the wall surface to $V_{0}$ at this
electrode. Possible trap designs are discussed in Sec. \ref{TrapDegign}. 

The side-wall
material can be beryllium, copper, or any other metal with a weak neutron absorption
and a rough surface to hold $^{4}$He film. We consider a one-dimensional
triangular roughness with period $l_{\mathrm{R}}\sim 10$~\textmu m and depth $h_{%
\mathrm{R}}\sim 1$~\textmu m, as proposed in Ref. \cite{Grigoriev2021} and
illustrated in Fig. \ref{FigTriang}. According to Ref. \cite{Grigoriev2021}, 
$^{4}$He film covering this wall is thick enough to protect UCN from the
absorption everywhere except the sharp triangular edges of the wall.
Fortunately, near these edges the electric field strength $E\left( 
\bm{r}\right) $ is much higher than the electric field $E_{0}\sim
V_{0}/L $ far from this edge, so that the last electric term in Eqs. (\ref%
{Ee}) or (\ref{EeSW}) is large enough to hold a sufficiently thick helium
film. To estimate the $^{4}$He film thickness near this corner we first need
to calculate the strength distribution of electric field. Since the $^{4}$He
dielectric constant $\varepsilon _{\mathrm{He}}=1.054$ is close to unity, we may
disregard the back influence of $^{4}$He on electric field, performing the
calculations for metal-vacuum interface.

 
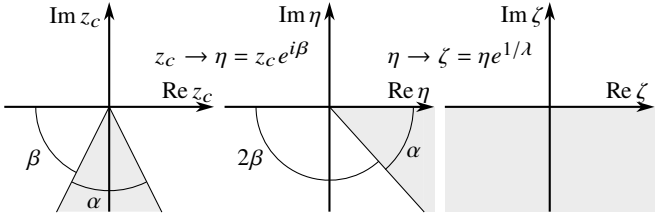
\begin{figure}[htb]
	\begin{tikzpicture}[every node/.style={inner sep=0,outer sep=0}]
		\node (picture) {\includegraphics[width=\linewidth]{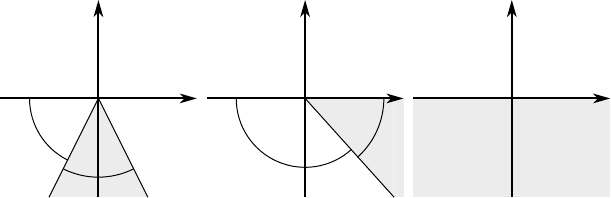}};
		
		\node[below=0.2cm,right=0.6cm] at (picture.north west) {$\operatorname{Im}z_c$};
		\node[below=1.2cm,right=2.1cm] at (picture.north west) {$\operatorname{Re}z_c$};
		
		\node[below=0.2cm,right=3.65cm] at (picture.north west) {$\operatorname{Im}\eta$};
		\node[below=1.2cm,right=5.1cm] at (picture.north west) {$\operatorname{Re}\eta$};
		
		\node[below=0.2cm,right=6.6cm] at (picture.north west) {$\operatorname{Im}\zeta$};
		\node[below=1.2cm,right=8cm] at (picture.north west) {$\operatorname{Re}\zeta$};
		
		\node[below=0.7cm,right=2cm] at (picture.north west) {$z_c \to \eta = z_c e^{i \beta}$};
		\node[below=0.7cm,right=5.1cm] at (picture.north west) {$\eta \to \zeta = \eta e^{1/\lambda}$};
		
		\node[above=0.7cm,right=0.3cm] at (picture.south west) {$\beta$};
		\node[above=0.13cm,right=1.1cm] at (picture.south west) {$\alpha$};
		
		\node[above=0.7cm,right=3.1cm] at (picture.south west) {$2\beta$};
		\node[above=0.8cm,right=5.35cm] at (picture.south west) {$\alpha$};
		
	\end{tikzpicture}
	
	\caption{Conformal mapping of infinitely long edge of angle $\protect\alpha =%
		\protect\pi -2\protect\beta $, used to find the solution of Laplace equation
		for electrostatic potential.}
	
	\label{FigConf}
\end{figure}

To estimate the electric field near an infinitely long edge of angle $\alpha
=\pi -2\beta $, $0<\beta <\pi /2$ (see Fig. \ref{FigConf}), we use the
standard method of conformal mapping of dimensionless complex coordinate $%
z_{c}\equiv (x+iy)/l=r\exp \left( i\phi \right) /l$ to $\zeta \equiv
u+iv=\rho \exp \left( i\theta \right) $ \cite{binns1973}: 
\begin{equation}
\zeta =\left( z_{c}e^{i\beta }\right) ^{1/\lambda }=r^{1/\lambda }\exp \left[
i\left( \phi +\beta \right) /\lambda \right] ,
\end{equation}%
where$~\lambda =1+2\beta /\pi $. This mapping transforms the angle to a line
and allows an easy calculation of the electric potential as 
\begin{equation}
V\left( \zeta \left( z\right) \right) =\Delta V_{0}\text{Im}\zeta =\Delta
V_{0}r^{1/\lambda }\sin \left[ \left( \phi +\beta \right) /\lambda \right] ,~
\label{V}
\end{equation}%
where $\Delta V_{0}$ is the potential raise per the normalization distance $%
l $. The electric field strength near the edge is
\begin{equation}
\left\vert E\right\vert =E_{0}\left\vert \frac{\mathrm{d}\zeta }{\mathrm{d}z_{c}}\right\vert =%
\frac{E_{0}}{\lambda }\left\vert z_{c}\right\vert ^{1/\lambda -1},  \label{E}
\end{equation}%
where $E_{0}=\Delta V_{0}/l$ is the electric field far from the edge. In our
case of periodic triangular wall roughness, shown in Fig. \ref{FigTriang},
the characteristic length scale $l=l_{\mathrm{R}}\sim 10$~\textmu m, because at a
distance $r\gg l_{\mathrm{R}}$ from the edge the electric field does not
have singularity and is almost uniform as near a flat wall. The electric
field squared at $r\ll l_{\mathrm{R}}$ 
\begin{equation}
E^{2}\left( \bm{r}\right) =\frac{E_{0}^{2}}{\lambda ^{2}}\left( \frac{l_{%
\mathrm{R}}}{r}\right) ^{4\beta /\left( \pi +2\beta \right) }.  \label{E2}
\end{equation}%
At a distance $r$ from the edge the electric field $E\left( r \right) $ is
larger than the average electric field $E_{0}\approx V_{0}/L$ by a factor of 
\begin{equation}
\gamma =E\left( r \right) /E_{0}=\left( l_{\mathrm{R} }/r\right) ^{2\beta
/\left( \pi +2\beta \right) }/\left( 1+2\beta /\pi \right) .  \label{EdHe}
\end{equation}%
We are interested to coat the edge of wall roughness with the $^{4}$He film
of thickness $d_{\mathrm{He}}^{\ast }\approx 100$~nm. At $r =d_{\mathrm{He}}^{\ast }=100$~nm, $%
l_{\mathrm{R}}=10$~\textmu m and $\beta =\pi /3$ this parameter $\gamma
=E\left( d_{\mathrm{He}}^{\ast }\right) /E_{0}\approx 3.8$, while at $\beta =\pi /4$
we obtain $\gamma =E\left( d_{\mathrm{He}}^{\ast }\right) /E_{0}\approx 3.1$.

In Eqs. (\ref{E2})-(\ref{EdHe}) $E\left( \bm{r}\right) \rightarrow
\infty $ at $\bm{r}=0 $ because we took an infinitely sharp edge. The
actual curvature radius at the edge $R_{c}$ is finite, and Eqs. (\ref{E})-(%
\ref{EdHe}) are valid at $r\gg R_{c}$. Hence, we may use Eqs. (\ref{E})-(\ref%
{EdHe}) at the $^{4}$He free surface if the curvature radius is much smaller
than the depth of $^{4}$He film at the edge, $R_{c}\ll d_{\mathrm{He}}$. In the
derivation of Eqs. (\ref{V})-(\ref{EdHe}) we considered a single edge. This
imposes another restriction on using Eqs. (\ref{E}),(\ref{E2}): $r\ll l_{%
\mathrm{R}}$. Thus, Eqs. (\ref{V})-(\ref{E2}) hold if 
\begin{equation}
R_{c}\ll r\approx d_{\mathrm{He}}^{\ast }\approx 100~\textrm{nm}\ll l_{\mathrm{R}}\sim 10~\textrm{\textmu m}.  \label{Cond}
\end{equation}%
The electric-field distribution $E\left( \bm{r}\right) $ near a periodic
boundary, such as a rectangular diffraction grating, can be studied using the
Fourier series and Rogowski's or Roth's methods (see ch. 5 of \cite%
{binns1973}), but it is much more complicated and gives a less visual result.

\begin{figure}[tbh]
	\begin{tikzpicture}[every node/.style={inner sep=0,outer sep=0}]
		\node (picture) {\includegraphics[width=\linewidth]{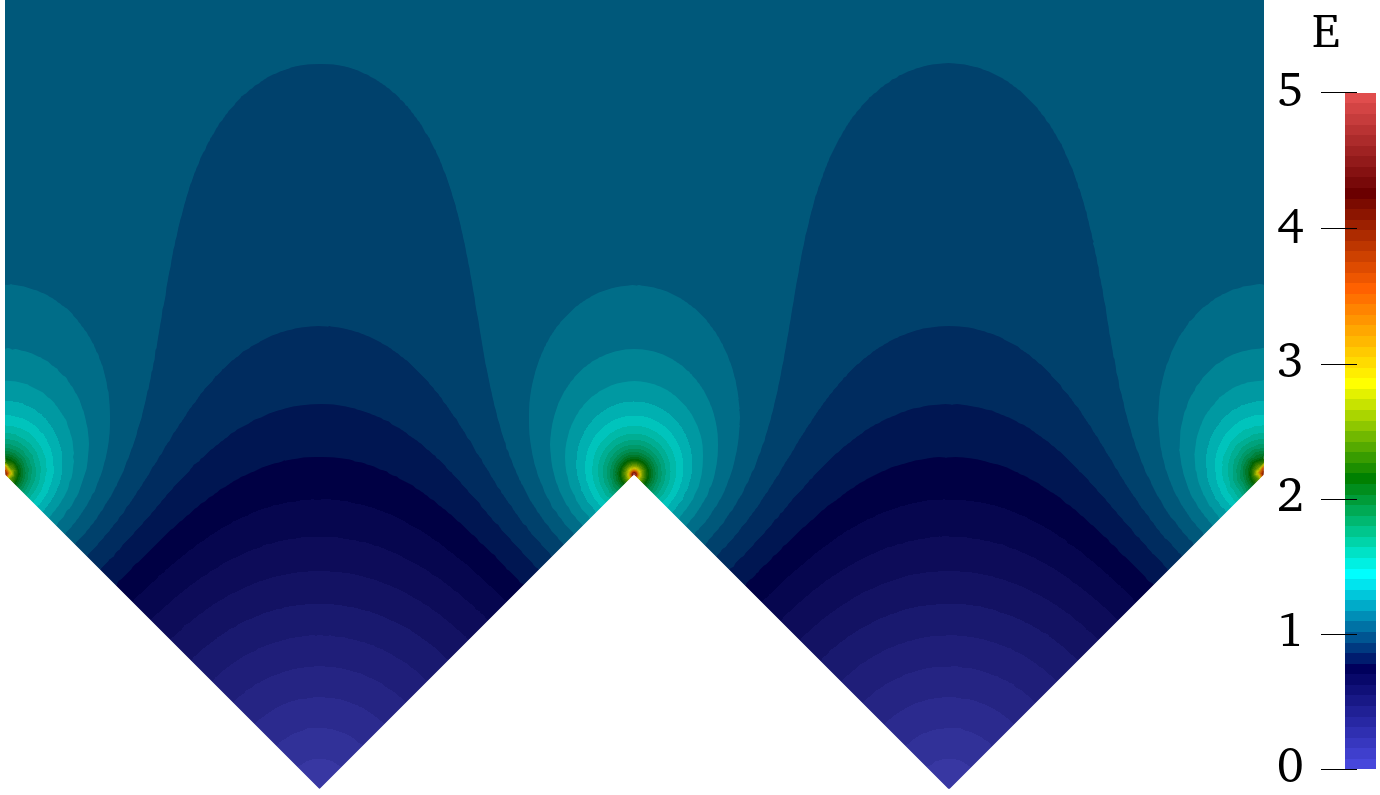}};
		\node[above right] at (picture.south west) {(a)};
	\end{tikzpicture}
	\\
	\begin{tikzpicture}[every node/.style={inner sep=0,outer sep=0}]
		\node (picture) {\includegraphics{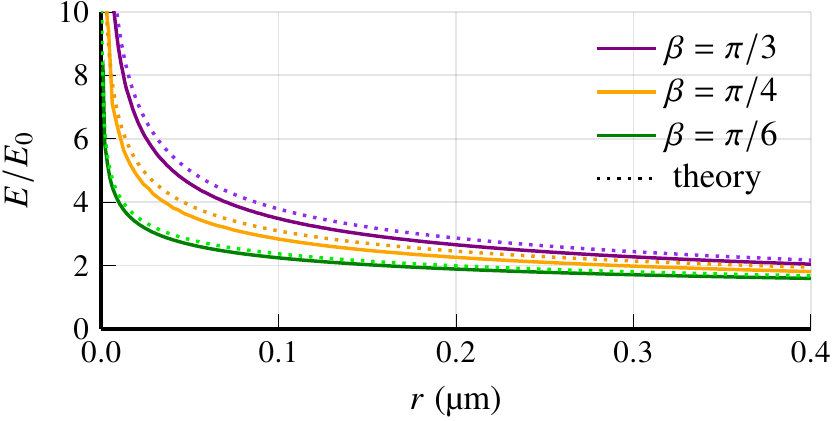}};
		\node[above right] at (picture.south west) {(b)};
	\end{tikzpicture}
	\caption{Numerical results for the electric field distribution $E(\mathrm{R})
		$ near a equipotential wall having the shape of a triangular grid, 
		i.e. a triangular groove periodically repeated along the wall. (a) Color map of
		the electric field strength. (b) Numerically calculated (solid curves)
		electric field strength $E$ as a function of the distance $r$ to the
		triangular wall edge in the direction $y$ perpendicular to the wall for
		several angles $\protect\beta$ of the triangles. It is compared to the
		analytical result given by Eq. (\protect\ref{EdHe}) and shown by dotted
		curves.}
	\label{FigNum}
\end{figure}

In our physical problem the condition (\ref{Cond}) is satisfied, but there
is another source of possible quantitative error -- the cutoff choice $l=l_{%
\mathrm{R}}$. This choice is only qualitatively correct, i.e. up to a factor 
$\sim 1$. To analyze the possible error we performed the numerical
calculation of the electric field distribution by solving the Laplace
equation for the electrostatic potential $V(\bm{r})$ using the method of
finite elements. The boundary conditions are taken as $V=0$ at the rough
trap wall of the periodic triangular shape, as shown in Fig. \ref%
{FigTriang}, and $V=V_0$ at the flat electrode parallel to this wall. Thus,
the numerical problem is 2D. The result of this calculation for the electric 
field distribution is given in Fig. \ref{FigNum}a,  and the
comparison of calculated parameter $\gamma \left( r \right) =E\left( r \right) /E_{0}$ 
with Eq. (\ref{EdHe}) is shown in Fig. \ref{FigNum}b for several edge angles $\alpha
=\pi -2\beta $.

The comparison of analytical formula in Eq. (\ref{EdHe}) and the
numerical result, shown in Fig. \ref{FigNum}, indicates that Eq. (\ref{EdHe}) 
describes very well the electric field distribution near the edge. Therefore, to
estimate the required electric field $E_0$ and voltage $V_0$ to hold $^{4}$%
He film of thickness $d_{\mathrm{He}}^{\ast }$ near the wall edge we use Eqs. (\ref%
{Emax})-(\ref{EdHe}) with the upper cutoff equal to the period $l_{\mathrm{R}%
}\sim 10$~\textmu m of the triangular grid. Eqs. (\ref{Emax}),(\ref{E2}) and (%
\ref{EdHe}) give the required strength $E_0$ of the quasi-uniform electric field
far from the wall edge: 
\begin{equation}
E_{0}\approx \frac{E^{\ast }}{\gamma} =\left( \frac{d_{\mathrm{He}}^{\ast }}{%
l_{\mathrm{R}}}\right) ^{2\beta /\left( \pi +2\beta \right) }\lambda \sqrt{%
\frac{4\pi \rho _{\mathrm{He}}gh}{\varepsilon _{\mathrm{He}}-1}}.  \label{eV}
\end{equation}%
For $d_{\mathrm{He}}=d_{\mathrm{He}}^{\ast }=100$~nm, $l_{\mathrm{R}}\sim
10$~\textmu m, and $\beta =\pi /4$, corresponding to straight edge angle $%
\alpha =\pi /2$, Eq. (\ref{eV}) gives $E_{0}=75$~\textrm{kV/cm}. Taking $%
\alpha =\beta =\pi /3$ slightly reduces the required electric field
intensity to $E_{0}\approx 60$~\textrm{kV/cm}.

The external electric field strength of $\sim 4$~\textrm{kV/cm} 
is rather common \cite{Volodin1977,Leiderer1992} in the experiments
with electrons on liquid helium surface, but raising this electric field by
an order of magnitude may be technically difficult. An electric field $>100$~\textrm{kV/cm} 
was experimentally realized in a $1$~cm gap between two
electropolished stainless steel electrodes $12$~cm in diameter for a wide
range of pressures at $T=0.4$~K \cite{Ito2016}. The effect of a weaker electric 
field $E_0\leq 45$~\textrm{kV/cm} on the superfluid helium
scintillation produced by fast electrons or by $\alpha $-particles at $T\geq
0.4$~K was also investigated experimentally \cite{Ito2012,Phan2020}.

\section{Effect of electric field on ripplon dispersion}

After solving the problem of coating the UCN trap walls by a sufficiently
thick helium film, which protects UCN from the absorption inside the trap walls, 
the most important factor limiting the precision
of UCN $\tau _{n}$ measurements is the inelastic neutron scattering by ripplons -- the quanta
of surface waves. At low temperature $T<0.5$~K, when the concentration of helium vapor is exponentially small, the main contribution to neutron scattering
rate comes from low-frequency ripplons with energy $\hbar \omega _{q}\sim
V_{0}^{He}=18.5\ \mathrm{neV}$ \cite{Grigoriev2016Aug}. The corresponding
ripplon wave vector is still much larger than the inverse capillary length $a_{\mathrm{He}}$ of $^{4} $He: 
\begin{equation}
	q_{0}\approx \left[ \left( V_{0}^{He}/\hbar \right) ^{2}\rho _{\mathrm{He}%
	}/\sigma _{\mathrm{He}}\right] ^{1/3}\approx 6.9~\textrm{\textmu m$^{-1}$}\gg \varkappa 
	\text{,}  \label{q0}
\end{equation}%
where $\varkappa =a_{\mathrm{He}}^{-1}=\sqrt{g\rho _{\mathrm{He}}/
	\sigma _{\mathrm{He}}}\approx 20$ cm$^{-1}$. Hence, the ripplon dispersion at this wave vector is given by that of capillary waves: $\omega _{q}=\sqrt{\sigma _{\mathrm{He}}/\rho _{\mathrm{He}}}\ q^{3/2}$. If the electric field increases the ripplon energy $\hbar \omega _{q}$, this reduces the equilibrium ripplon density and the UCN scattering rate by ripplons.

\subsection{Uniform electric field}

In a uniform electric field the ripplon dispersion law modifies \cite%
{Melnikovskii1997} to%
\begin{equation}
	\omega _{q}^{2}=gq+\frac{\sigma _{\mathrm{He}}}{\rho _{\mathrm{He}}}q^{3}+%
	\frac{\left( \varepsilon -1\right) ^{2}}{4\pi \rho _{\mathrm{He}}\varepsilon
		\left( \varepsilon +1\right) }\left( \varepsilon E_{||}^{2}\cos ^{2}\theta
	-E_{\perp }^{2}\right) q^{2},  \label{2}
\end{equation}%
where $\varepsilon =\varepsilon _{\mathrm{He}}=1.054$ and the angle $\theta $
is between the electric field and the ripplon wave vector. For an electric
field parallel to helium surface, as in Fig. \ref{FigTrap}, the
field-induced correction to ripplon dispersion is positive and have a lower
power of wave vector than the dominating capillary term. If this correction
is large enough, it may reduce the UCN scattering rate by ripplons. The
ratio of the last electric term in Eq. (\ref{2}), arising from $^{4}$He
polarization, to the second term, coming from capillary effect, for an
electric field along the surface and parallel to $q$-vector, $E_{\perp }=0$,
and $\theta =0$, is 
\begin{equation}
	\nu \equiv \frac{\left( \varepsilon -1\right) ^{2}E_{||}^{2}}{4\pi \left(
		\varepsilon +1\right) \sigma _{\mathrm{He}}}\frac{1}{q_{0}}=\frac{\left(
		\varepsilon -1\right) ^{2}e^{2}E_{||}^{2}}{4\pi \left( \varepsilon +1\right)
		\sigma _{\mathrm{He}}}\frac{1}{e^{2}q_{0}}.
\end{equation}%
Unfortunately, for $^{4}$He in a reasonably strong external electric field $%
E_{||}=E_{0}=10$~kV/cm and at $q=q_{0}$ this ratio is too small: $%
\nu \approx 5 \cdot 10^{-6}$. Even at $E_{||}=E_{\ast }=230$~\textrm{kV/cm}
this ratio at $q=q_{0}$ is not sufficient to change the ripplon 
dispersion considerably: $\nu \left( E_{\ast }\right) \approx
2.6\cdot 10^{-3}\ll 1$. Hence, the correction to ripplon dispersion from a $%
^{4}$He polarization in a uniform electric field, given by the last term in
Eq. (\ref{2}), is negligibly small and cannot help to reduce the UCN
scattering rate by ripplons.

\subsection{Nonuniform electric field.}

Eq. (\ref{2}) is derived for a uniform electric field. A high nonuniform
electric field, as on the surface of thin helium film near the edges of wall
roughness, may change the ripplon dispersion stronger. According to Eq. (\ref%
{Ee}), a nonuniform electric field creates a force%
\begin{equation}
	\bm{F} = -\nabla E_{\mathrm{e}}=\frac{\varepsilon _{He}-1}{4\pi }\int \!\mathrm{d}^{3}\bm{r}
	\nabla \left[ E^{2}\left( \bm{r}\right) \right]   \label{Fe}
\end{equation}%
acting on $^{4}$He volume. This force is added to the gravity force $F_{\mathrm{g}}=\rho _{\mathrm{He}} \bm{g} \int \!\mathrm{d}^{3}\bm{r}$ and renormalizes the
free fall acceleration as%
\begin{equation}
	\bm{g} \rightarrow \bm{g}^{\ast }\left( \bm{r}\right) = \bm{g}-\frac{\varepsilon _{He}-1}{%
		4\pi \rho _{\mathrm{He}}}\nabla \left[ E^{2}\left( \bm{r}\right) \right]
	.  \label{gr}
\end{equation}%
For the triangular side-wall roughness near a sharp edge, $\beta \rightarrow
\pi /2$, covered by $^{4}$He film of thickness $d_{\mathrm{He}}^{\ast }=100$~nm we have 
\begin{equation}
	g^{\ast }\left( d_{\mathrm{He}}^{\ast }\right) \sim \frac{\varepsilon
		_{He}-1}{4\pi \rho _{\mathrm{He}}}\frac{E_{\ast }^{2}}{d_{\mathrm{He}}^{\ast
	}}\approx 1.75\cdot 10^{6}g.
\end{equation}%

\begin{figure}[htb]
	\includegraphics{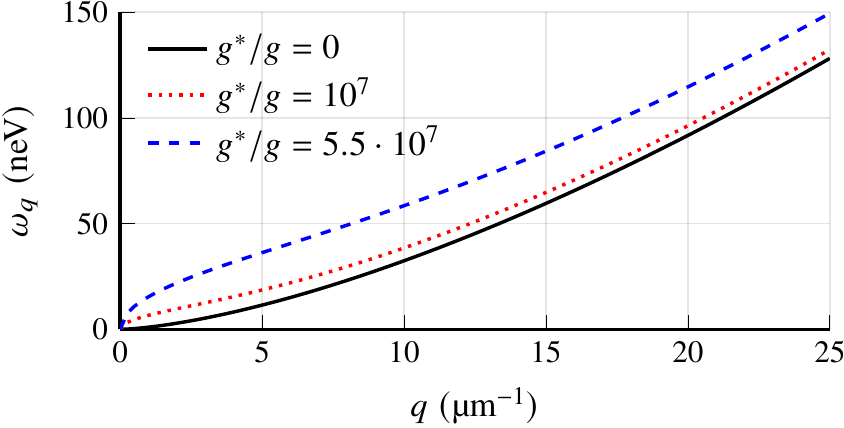}
	\caption{Ripplon dispersion, given by Eq. (\ref{2m}).}
	\label{FigDisp}
\end{figure}

To estimate the effect of nonuniform electric field near the wall edge
covered by $^{4}$He film at arbitrary $\beta $ we substitute Eq. (\ref{E2})
to Eq. (\ref{gr}), which gives: 
\begin{equation}
	\frac{g^{\ast }\left( \bm{r}\right) }{g}=1+\frac{\varepsilon _{He}-1}{%
		4\pi \rho _{\mathrm{He}}g}\frac{4\beta \pi ^{2}E_{0}^{2}}{r\left( \pi
		+2\beta \right) ^{3}}\left( \frac{l}{r}\right) ^{\frac{4\beta }{\pi +2\beta }%
	}.  \label{gEdge}
\end{equation}%
For $r=d_{\mathrm{He}}^{\ast }=100$~nm, $l=l_{\mathrm{R}}=10$~\textmu m, $%
E_{0}=60$~\textrm{kV/cm}, and $\beta =\pi /3$ this gives $g^{\ast }/g\approx
1.4\cdot 10^{6}$. The corresponding $\varkappa ^{\ast }=
\sqrt{g^{\ast }\rho_{\mathrm{He}}/\sigma _{\mathrm{He}}}\approx 2.4$~\textmu m$^{-1}$ $\sim q_{0}$.
Hence, the ripplon dispersion changes considerably at $q=q_{0}$ due to such
a nonuniform electric field. Since at the trap bottom of UCN trap we do not
need to hold liquid helium by a capillary effect, we may take a larger $l=l_{%
	\mathrm{R}}\sim 1$~mm. Then, according to Eq. (\ref{gEdge}) this
raises $g^{\ast }/g$ about $100^{4/5}\approx 40$ times to $g^{\ast
}/g\approx 5.5\cdot 10^{7}$. The corresponding ripplon dispersion, given by
Eq. (\ref{2}) without the last term but with renormalized $g^{\ast }$,%
\begin{equation}
	\hbar \omega _{q}=\hbar \sqrt{g^{\ast }q+\frac{\sigma _{\mathrm{He}}}{\rho _{%
				\mathrm{He}}}q^{3}},  \label{2m}
\end{equation}%
is shown in Fig. \ref{FigDisp}. It illustrates a considerable increase in ripplon energy $\hbar \omega _{q}$ at $q=q_{0}$. Hence, theoretically, a nonuniform electric
field may reduce the UCN inelastic scattering rate by thermally activated
ripplons. However, a more thorough calculation is needed to study this effect quantitatively at
a nonuniform gradient of the electric field strength.

\section{Possible trap designs}\label{TrapDegign}

\begin{figure}[htb]
	\begin{tikzpicture}[every node/.style={inner sep=0,outer sep=0}]
		\node (picture) {\includegraphics[width=0.9\linewidth]{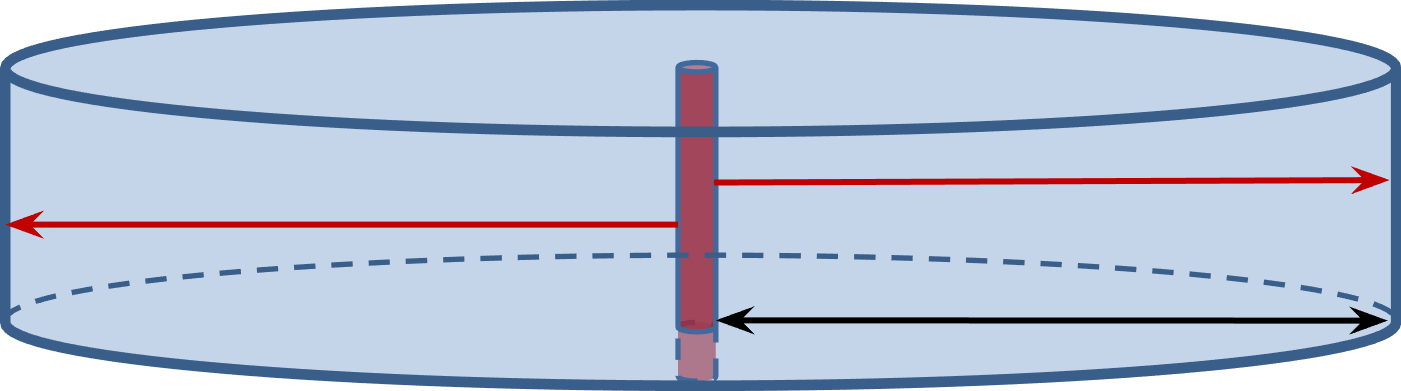}};
		\node[above=0cm,right] at (picture.south west) {(a)};
		
		\node[below=0cm,right=0.2cm] at (picture.north west) {$V=0$};
		\node[below=0.5cm,right=4.0cm] at (picture.north west) {$V=V_0$};
		\node[below=1cm,right=0.2cm] at (picture.north west) {$E$};
		\node[below=1.63cm,right=5.7cm] at (picture.north west) {$L$};
	\end{tikzpicture}
	\begin{tikzpicture}[every node/.style={inner sep=0,outer sep=0}]
		\node (picture) {\includegraphics[width=0.9\linewidth]{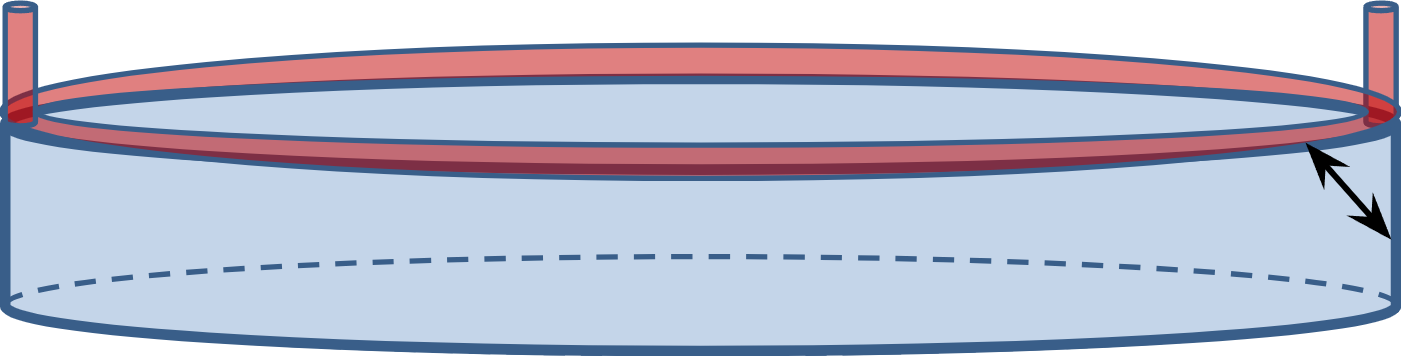}};
		\node[above=-0.1cm,right] at (picture.south west) {(b)};
		
		\node[below=0.2cm,right=0.25cm] at (picture.north west) {$V=V_0$};
		\node[below=1.2cm,right=0.2cm] at (picture.north west) {$V=0$};
		\node[below=1.25cm,left=0.3cm] at (picture.north east) {$L$};
	\end{tikzpicture}
	\begin{tikzpicture}[every node/.style={inner sep=0,outer sep=0}]
		\node (picture) {\includegraphics[width=0.9\linewidth]{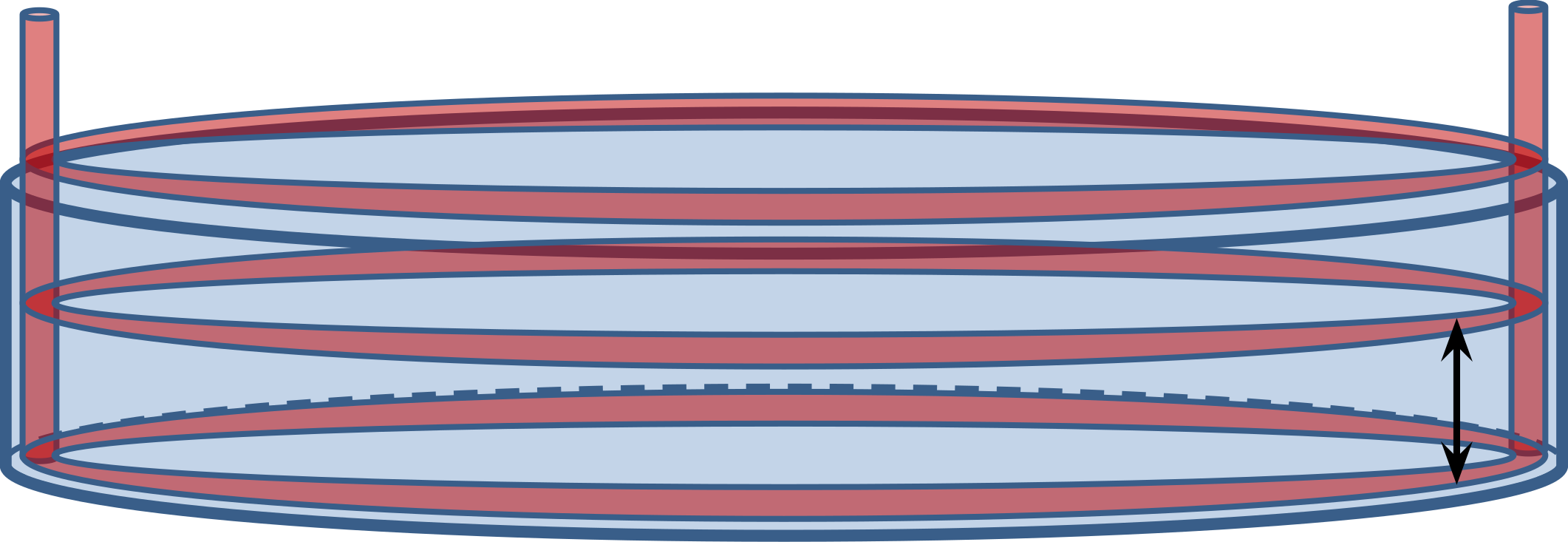}};
		\node[above=0cm,right] at (picture.south west) {(c)};
		
		\node[below=0.4cm,right=0.3cm] at (picture.north west) {$V=V_0$};
		\node[below=0.0cm,left=0.0cm] at (picture.south east) {$V=0$};
		\node[below=1.9cm,left=0.6cm] at (picture.north east) {$L$};
	\end{tikzpicture}
	\begin{tikzpicture}[every node/.style={inner sep=0,outer sep=0}]
		\node (picture) {\includegraphics[width=0.9\linewidth]{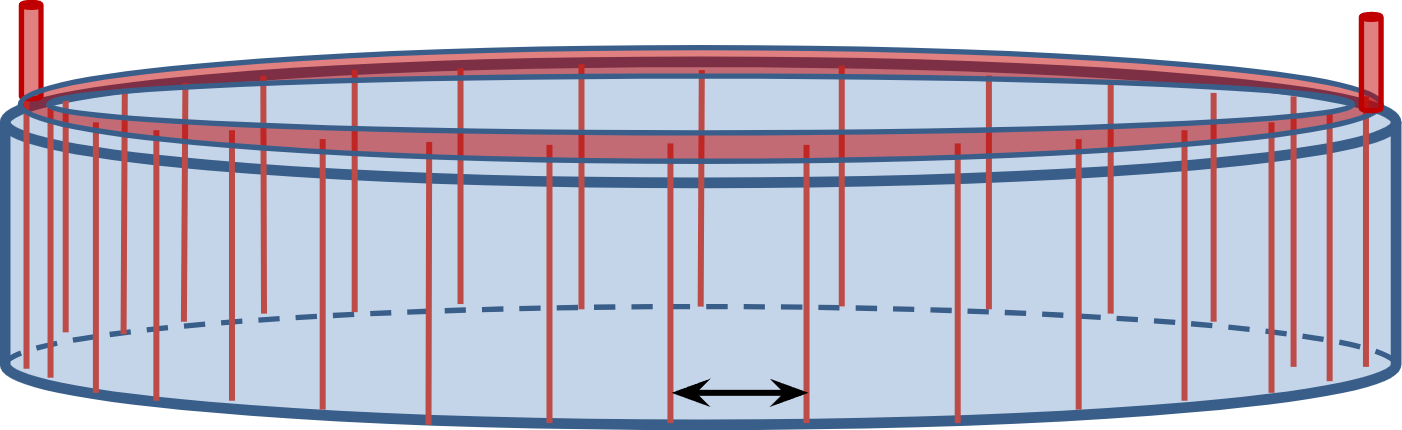}};
		\node[above=-0.05cm,right] at (picture.south west) {(d)};
		
		\node[below=0.2cm,right=0.25cm] at (picture.north west) {$V=V_0$};
		\node[below=0.05cm,left=0.0cm] at (picture.south east) {$V=0$};
		\node[below=2cm,right=4.0cm] at (picture.north west) {$L$};
	\end{tikzpicture}
	
	\caption{Possible UCN trap designs with electrostatic potential.}
	\label{FigTrap}
\end{figure}
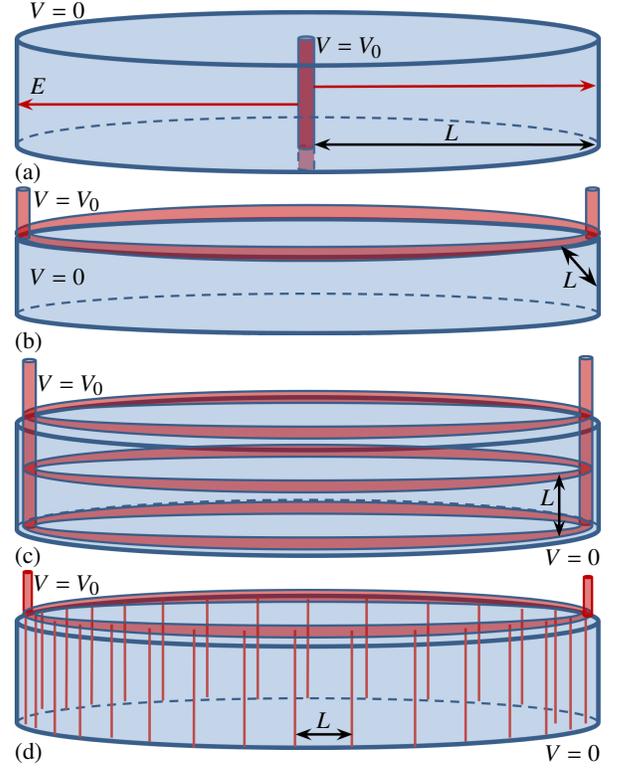

The voltage $V_0$, corresponding to the required electric field $E_0$ and $%
E_*$, depends on the geometry of electrodes. In the trap design of Fig. \ref%
{FigTrap}a with the trap radius $L=R_{\mathrm{trap}}=1~\mathrm{m}$ a strong field $%
E_{0}\approx 60$~\textrm{kV/cm} requires a voltage difference $%
V_{0}=E_{0}L\approx 6~\mathrm{MV}$, which is too high. In the trap design
shown in Figs. \ref{FigTrap}b,c the electric field $E_{0}$ and voltage $V_{0}
$ do not depend on the trap radius $R_{\mathrm{trap}}$ but only on the distance from
the grounded side wall to the electrode at $V=V_{0}$.

The trap in Figs. \ref{FigTrap}b has only one toroidal electrode at $V=V_{0}$
placed at height $h_{\max }=V_{0}^{\mathrm{He}}/m_{n}g\approx 18$cm above the helium
level, i.e. just above the maximal height of UCN inside the trap. Then the
distance $L\approx h_{\max }-h$ depends on the height $h$ on the wall from
the helium level. At large height $h\sim h_{\max }$, where the gravity
energy to be compensated by electric field is the highest, the electric
field $E_{0}\left( h\right) =V_{0}/L\approx V_{0}/\left( h_{\max }-h\right) $
is also the highest. Then Eq. (\ref{eV}) gives 
\begin{equation}
V_{0}=\left( \frac{d_{\mathrm{He}}}{l_{\mathrm{R}}}\right) ^{2\beta /\left(
\pi +2\beta \right) }\lambda \sqrt{\frac{4\pi \rho _{\mathrm{He}}gh}{\varepsilon
_{\mathrm{He}}-1}}\left( h_{\max }-h\right) ,
\end{equation}%
which has a maximal value 
\begin{equation}
V_{0}^{\max }=0.385\left( \frac{d_{\mathrm{He}}}{l_{\mathrm{R}}}\right)
^{2\beta /\left( \pi +2\beta \right) }\lambda \sqrt{\frac{4\pi \rho
_{\mathrm{He}}gh_{\max }^{3}}{\varepsilon _{\mathrm{He}}-1}},  \label{V0m}
\end{equation}%
at $h=h_{\max }/3\approx 6$~cm. At this heigh one can take \cite%
{Grigoriev2021} $l_{\mathrm{R}}=4a_{\mathrm{He}}^{2}/h\approx 17$~\textmu m. Substituting
this and $\alpha =\beta =\pi /3$ to Eq. (\ref{V0m}) we obtain $V_{0}^{\max
}\approx 350$~\textrm{kV} $\approx E_{0}h_{\max }/3$, which is still very high
and technically difficult. In Ref. \cite{Ito2016} the realized voltage
difference between two electrodes in $^{4}$He was only $V_{0}=100$~\textrm{kV}.

The required voltage $V_{0}$ can be reduced by an order of magnitude or more if one
uses the trap design shown in Fig. \ref{FigTrap}c with several toroidal
electrodes at $V=V_{0}$ placed at different heights $h_{i}<h_{\max } $ above
the helium level. If these toroidal electrodes are rather thin and placed at
a small distance $\sim $ $l_{\mathrm{R}}$ from the trap wall, theoretically,
one can reduce the required voltage to only $V_{0}\sim E_{0}l_{\mathrm{R}%
}\sim 60$~\textrm{V}. Technically,
it may be more convenient to use the trap design illustrated in Fig. \ref%
{FigTrap}d, where thin wire electrodes hang down from the toroidal
electrode above $h=h_{\max }$. 

An electrode at height $h<h_{\max }$ may produce additional inelastic
scattering of UCN inside the trap. This electrode is always covered by a
helium film of thickness $d>10~\mathrm{nm}$ due to the van-der-Waals forces,
which attract helium vapor. However, for a save protection of UCN by this $%
^{4}$He film covering the electrodes we need $d\geq d_{\mathrm{He}}^{\ast
}=100~\mathrm{nm}$. Such a thick $^{4}$He film can be hold by either the
surface roughness and capillary effect \cite{GrigorievPRC2021,Grigoriev2021}%
, described by Eq. (\ref{eq:Es}), or by the electrostatic energy of $^{4}$He
in electric field, given by Eq. (\ref{Ee}). The latter is sufficient for a
rather thin electrode. Indeed, the electric field around a cylindrical
electrode of radius $R_{e}$ is $E\left( \bm{r}\right) \approx
V_{0}/r/\ln \left( L/R_{e}\right) \approx E_{0}L/r$. Hence, according to
Eqs. (\ref{EdHe}) and (\ref{eV}), a commercially available copper wire of radius $%
R_{e}=10$~\textmu m \footnote{Bobbins with ultra thin copper wires of diameter $10$~\textmu m are
available for $\$10.99$ at www.amazon.com.} at
voltage $V_{0}$ placed at a distance $L=200$~\textmu m = 0.2 mm from the grounded
wall holds a $^{4}$He film of sufficient thickness $d_{\mathrm{He}}^{\ast
}=100~\mathrm{nm}\ll R_{e}$ if 
\begin{equation*}
E\left( d_{\mathrm{He}}^{\ast }\right) \approx \frac{V_{0}}{R_{e}\ln \left(
L/R_{e}\right) } \geq E_{\ast }\approx 230~\mathrm{kV/cm},
\end{equation*}%
or%
\begin{equation}
V_{0}\geq V_{0}^{w}=E_{\ast }R_{e}\ln \left( L/R_{e}\right) 
.  \label{Vow}
\end{equation}%
For $R_{e}=10$~\textmu m and $L=0.5$~mm this gives 
$V_{0}\geq 900~\mathrm{V}$. Thus, a sufficient $^4$He coating of thin electrodes inside UCN trap is easy, 
because the voltage $V_0$ required for this coating is smaller 
than to hold $^4$He film of sufficient thickness $d_{\mathrm{He}}^{\ast }=100~\mathrm{nm}$ 
on the side wall at height $h_{\max}=18$~cm above helium level. The latter at 
$\alpha =\beta =\pi /3$ requires $E_{0}\approx 60$~kV/cm, 
which for the $L=0.2$~mm gives $V_{0}\approx 1.2$~kV. Hence,
theoretically, by using the trap design in Figs. \ref{FigTrap}c,d one may
reduce the required voltage to only $V_{0}\sim 1$~kV. 
 Keeping two electrodes with
voltage difference $V_{0}\sim 1$~kV at a distance of only $0.2$~mm
remains technically difficult. One can increase the distance between electrodes 
at the cost of increasing their voltage difference. A smaller voltage holds a thinner 
$^{4}$He film near the wall roughness edges, but it may still be useful to protect
UCN from the absorption inside the trap walls. 

\begin{figure}[tbh]
	\includegraphics[width=\linewidth]{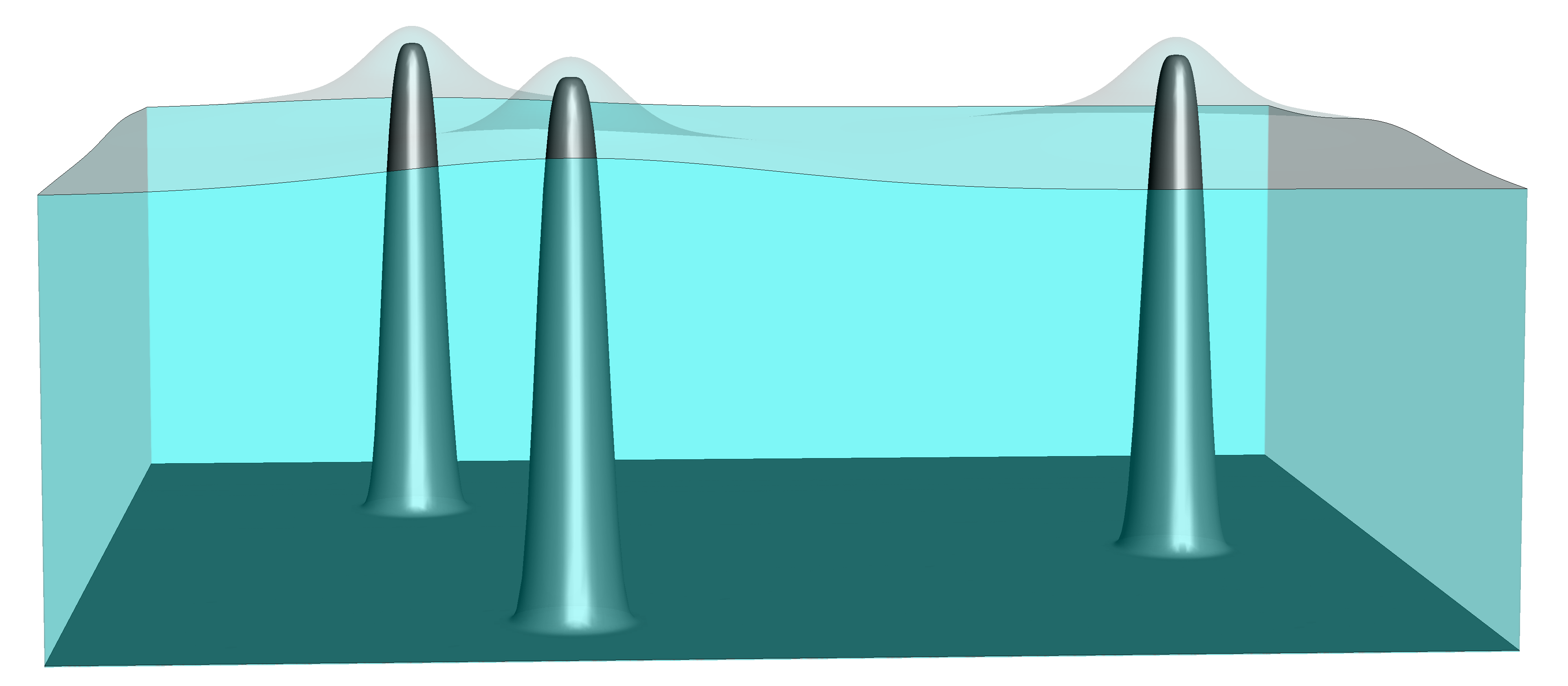}
	\caption{Schematic illustration of a 3D fur-like roughness of a metallic UCN trap wall holding a liquid helium film (light blue filling) by capillary effects. The applied electric field is the strongest near the wall peaks and holds additional amount of $^4$He around these peaks, providing a complete coating of the UCN trap wall.}
	\label{FigFur}
\end{figure}

A 3D fur-like wall roughness with
pyramidal or needle-shaped protrusions, as illustrated in Fig. \ref{FigFur}, may further reduce the required voltage $V_{0}$. The electric field at a distance $r$ from a needle-like electrode at voltage $U$ of curvature radius $r_e$ above another electrode in the form of a plane perpendicular to the needle and separated by the distance $L$ is \cite{Florkowska1993} 
\begin{equation}
E(r)\approx \frac{2U/\ln (4L/r_e)}{ 2r+r_e-r^2/L}.
 \label{ETip}
\end{equation}%
Hence, in our case $L\gg l_{\mathrm{R}} \gg  r\sim d_{\mathrm{He}}^{\ast
}>r_e$ we have
\begin{equation}
	E(r)\approx \frac{E_{\ast }d_{\mathrm{He}}^{\ast
	}}{r+r_e/2} \approx  \frac{E_0 l_{\mathrm{R}}}{r+r_e/2}.
	\label{ETip1}
\end{equation}%
For $l_{\mathrm{R}}\gtrsim 5$~\textmu m and $r= d_{\mathrm{He}}^{\ast
}=100$~ nm, the required electric field at the surface $E_{\ast }\approx 230$~kV/cm corresponds to the electric field $E_{0 }\lesssim 5~\mathrm{kV/cm}$ far from the edge. If the second electrode is separated by a distance $L= 1$ mm from the wall, the required voltage is $V_0=500$ V.  
However, making of such a wall with 3D fur-like roughness is more difficult than a cheap 
triangular diffraction grating, studied above.

\section{Conclusions}

To summarize, we propose to improve the material UCN traps by coating with liquid helium using the combined effect of capillarity and of electric field. 
The side wall roughness holds a sufficiently thick $^4$He film by the capillary effects \cite{GrigorievPRC2021,Grigoriev2021}, 
but the very edges of this roughness remain coated by a too thin $^4$He film. If this rough wall serves as an electrode, the 
electric field is the strongest near these wall edges, which attracts $^4$He due to polarization forces. 
This helps to cover the wall edges and its entire surface by a sufficiently thick $^4$He film to completely protect UCN from the absorption inside trap walls. The second electrode, if made in the form of thin wires, may be placed inside the UCN trap because it get also coated by $^4$He and does not absorb neutrons. The strong nonuniform electric field on the helium surface increases the ripplon energy, which makes their equilibrium concentration smaller. This reduces the inelastic scattering rate of UCN by ripplons, but the effect is not sufficient to destroy this channel of UCN losses, which becomes dominating after the absorption of UCN inside trap walls is eliminated by their coating with liquid $^4$He. Fortunately, the neutron-ripplon interaction is weak, and the linear temperature dependence of UCN scattering rate by ripplons helps to accurately take this systematic error into account \cite{Grigoriev2016Aug}. A low temperature $T<0.5$ K of trap walls is required to eliminate another source of UCN losses -- scattering by helium vapor. 

In spite of the mentioned technical difficulties,
the proposed complete coating of UCN trap walls by liquid $^4$He may give rise to a new generation of ultracold neutron traps with a very long storage time. This may strongly improve the precision of neutron lifetime measurements and of other experiments with ultracold neutrons.  

\section{Acknowledgements}

The work of P.D.G. is supported by the Russian Science Foundation grant \# 23-22-00312. A.V.S. thanks the Foundation for the Advancement of Theoretical Physics and Mathematics ''Basis'' for grant \# 22-1-1-24-1. V.D.K. acknowledges the financial support from the NUST "MISIS" grant No. K2-2022-025 in the framework of the federal academic leadership program Priority 2030. A.M.D. acknowledges the Ministry of Science and Higher Education of the Russian Federation (state assignment no. 0033-2019-0001 “Development of the Theory of Condensed Matter”).

\bibliographystyle{apsrev4-2}
\input{UCN2023.bbl}

\end{document}

%% file: UCN2023.bbl
%